\documentstyle[12pt]{article}

\makeatletter

\def\@authoraddress{}
\def\@title{}
\def\title#1{\gdef\@title{{\par\vskip-10pt\Large\bf
\baselineskip20pt\centering\ignorespaces\uppercase{#1}\vskip6pt}}%
\setcounter{table}{0}      \setcounter{figure}{0}
\setcounter{equation}{0}   \setcounter{section}{0}
\setcounter{subsection}{0} \setcounter{subsubsection}{0}
\setcounter{paragraph}{0}
}

\def\authors#1{\expandafter\def\expandafter\@authoraddress\expandafter
{\@authoraddress %
{\dimen0=-\prevdepth \advance\dimen0 by1.5\baselineskip
\nointerlineskip \centering
\vrule height\dimen0 width0pt\relax\ignorespaces\large\sc#1\par
}%
}%
}

\def\addresses#1{\expandafter\def\expandafter\@authoraddress\expandafter
{\@authoraddress{\nointerlineskip\vskip1pc
                 \footnotesize\it\centering\ignorespaces#1\par}}}

\def\@maketitle{%
\@title
\ifdim\prevdepth=-1000pt \prevdepth0pt\fi
\@authoraddress
}

\def\maketitle{\par
\begingroup
\let\cite\@bylinecite
\global\@topnum\z@ %
\@maketitle
\endgroup
\def\@thanks{}\def\@authoraddress{}\def\@title{}
}

\def\abstract{\par
\bgroup
\ifdim\prevdepth=-1000pt \prevdepth0pt\fi
\hsize\columnwidth
\leftskip=2em \rightskip\leftskip
\dimen0=-\prevdepth \advance\dimen0 by2pc \nointerlineskip
\noindent\vskip1.5\baselineskip\nointerlineskip\noindent\footnotesize\relax}

\newif\if@firststuff

\def\endabstract{\par
\nointerlineskip \vskip0pt
\noindent \par
\egroup
\hrule depth0pt width0pt
\global\everypar{\global\@firststufffalse}\global\@firststufftrue
}

\renewcommand\subsection{\@startsection{subsection}{2}{\z@}%
                                     {-3.25ex\@plus -1ex \@minus -.2ex}%
                                     {1.5ex \@plus .2ex}%
                                     {\normalfont\large\bfseries}}

\def\1ad{\mbox{\normalsize $^1$}}
\def\2ad{\mbox{\normalsize $^2$}}
\def\3ad{\mbox{\normalsize $^3$}}
\def\4ad{\mbox{\normalsize $^4$}}
\def\5ad{\mbox{\normalsize $^5$}}
\def\6ad{\mbox{\normalsize $^6$}}
\def\7ad{\mbox{\normalsize $^7$}}
\def\8ad{\mbox{\normalsize $^8$}}
\def\adref#1{\mbox{\normalsize $^{#1}$}}
\makeatother

\begin{document}


\title{Progress in Higher Spin Gauge Theories}


\authors{M.A.Vasiliev\adref{1}}


\addresses{\1ad
I.E.Tamm Department of Theoretical Physics, Lebedev Physical Institute,\\
Leninsky prospect 53, 117924, Moscow, Russia}


\maketitle


\begin{abstract}
General properties of the theory of higher spin gauge
fields in $AdS_4$ are discussed. Some new results on the 4d conformal
higher spin gauge theory are announced. Talk given at the International
Conference "Quantization, Gauge Theory and Strings" in memory of
E.S.Fradkin, Moscow, June 5-10, 2000.
\end{abstract}



\def\mco{\multicolumn}
\def\epp{\epsilon^{\prime}}
\def\vep{\varepsilon}
\def\ra{\rightarrow}
\def\ppg{\pi^+\pi^-\gamma}
\def\vp{{\bf p}}
\def\ko{K^0}
\def\kb{\bar{K^0}}
\def\al{\alpha}
\def\ab{\bar{\alpha}}
\def\be{\begin{equation}}
\def\ee{\end{equation}}
\def\bea{\begin{eqnarray}}
\def\eea{\end{eqnarray}}
\newcommand{\bee}{\begin{eqnarray}}
\newcommand{\eee}{\end{eqnarray}}
\newcommand{\nn}{\nonumber}
\newcommand{\hy}{\hat{y}}
\newcommand{\by}{\bar{y}}
\newcommand{\bz}{\bar{z}}
\newcommand{\go}{\omega}
\newcommand{\Go}{\Omega}
\newcommand{\Gl}{\Lambda}
\newcommand{\Gpo}{{\Omega^\prime}}
\newcommand{\hsa}{$hsc^\infty (4)\,\,$}
\newcommand{\e}{\epsilon}
\newcommand{\half}{\frac{1}{2}}
\newcommand{\ga}{\alpha}
\newcommand{\gal}{\alpha}
\newcommand{\U}{\Upsilon}
\newcommand{\ups}{\upsilon}
\newcommand{\bu}{\bar{\upsilon}}
\newcommand{\dga}{{\dot{\alpha}}}
\newcommand{\dgb}{{\dot{\beta}}}
\newcommand{\gb}{\beta}
\newcommand{\gga}{\gamma}
\newcommand{\gd}{\delta}
\newcommand{\gl}{\lambda}
\newcommand{\gk}{\kappa}
\newcommand{\gep}{\epsilon}
\newcommand{\gvep}{\varepsilon}
\newcommand{\gs}{\sigma}
\newcommand{\V}{|0\rangle}
\newcommand{\ws}{\wedge\star\,}
\newcommand{\gee}{\epsilon}
\newcommand{\ggg}{\gamma}
\newcommand\ul{\underline}
\newcommand\un{{\underline{n}}}
\newcommand\ull{{\underline{l}}}
\newcommand\um{{\underline{m}}}
\newcommand\ur{{\underline{r}}}
\newcommand\us{{\underline{s}}}
\newcommand\up{{\underline{p}}}
\newcommand\uq{{\underline{q}}}
\newcommand\ri{{\cal R}}
\newcommand\punc{\multiput(134,25)(15,0){5}{\line(1,0){3}}}
\newcommand\runc{\multiput(149,40)(15,0){4}{\line(1,0){3}}}
\newcommand\tunc{\multiput(164,55)(15,0){3}{\line(1,0){3}}}
\newcommand\yunc{\multiput(179,70)(15,0){2}{\line(1,0){3}}}
\newcommand\uunc{\multiput(194,85)(15,0){1}{\line(1,0){3}}}
\newcommand\aunc{\multiput(-75,15)(0,15){1}{\line(0,1){3}}}
\newcommand\sunc{\multiput(-60,15)(0,15){2}{\line(0,1){3}}}
\newcommand\dunc{\multiput(-45,15)(0,15){3}{\line(0,1){3}}}
\newcommand\func{\multiput(-30,15)(0,15){4}{\line(0,1){3}}}
\newcommand\gunc{\multiput(-15,15)(0,15){5}{\line(0,1){3}}}
\newcommand\hunc{\multiput(0,15)(0,15){6}{\line(0,1){3}}}
\newcommand\ls{\!\!\!\!\!\!\!}

\section{Introduction}
\label{Introduction}

Higher spin theory is a classical subject of the field theory. It
takes origin from the  works of Dirac \cite{Dir},
Fierz and Pauli \cite{FP}, Rarita and Schwinger \cite{RS}
and many others. It is also
a traditional subject in the Theory Department of the Lebedev
Physical Institute where Professor Efim Samoilovich
Fradkin was working most of his scientific life and
where this problematics was initiated by Tamm and Ginzburg \cite{DHS}
(see \cite{G} for more historical remarks and references).
In fact, Fradkin got a position at the Theory Department
by recommendation of Vitaly Lazarevich Ginzburg who liked
the investigation of Efim Samoilovich on the theory of spin 5/2 field,
the subject of his first scientific paper  \cite{Fr}.

The concept of higher spins is time-dependent. For about
forty years the main interest was focused on the massive fields and
``higher spin" (HS) meant $s\geq 3/2$ (sometimes, even $s=1$). The situation
changed after creation of supergravity in 1976
\cite{FFN,DZ}. One possible interpretation of supergravity is that
it results from  resolution of the problem of consistent interactions
for the massless spin 3/2 field. A lesson  was that the
principle of gauge symmetry associated with the corresponding massless
field (local supersymmetry for $s=3/2$) played a key role in the theory.
As a result, since 1976 the interest shifted to the massless fields
associated with gauge symmetries
with the updated convention that ``higher spin" implies $s\geq 5/2$.

In the early seventies Efim Samoilovich was mainly interested in
the HS theory as a range for the application of the new
methods of quantization of constrained systems he was working on.
As an undergraduate student of Efim Samoilovich I was lucky to get
a problem related to the Hamiltonian analysis of the spin 3/2 and
spin 2 massive fields in external fields thus learning both the
constrained dynamics and the HS theory. However,
one day in spring of
1976 my research plans have changed drastically after I got a phone
call from Efim Samoilovich with the urgent instruction to go to the
Moscow international airport Sheremet'evo to get
a copy of some new important paper from Vladimir Kadyshevsky
coming from CERN. Needless to say that
those days it was the most efficient way of mailing news to USSR.
What I got was the first paper on supergravity by Ferrara, Freedman
and van Nieuwenhuizen \cite{FFN}. This case illustrates
perfectly how precise was the feeling of Efim Samoilovich of what
is really important in science.

Supergravity became one of the main scientific interests of
Efim Samoilovich. A development having particular importance for
the HS problem was N=2 gauged supergravity
found independently be Freedman and Das \cite{DF} and by
us \cite{fvN2}. What we learned from this model was that
at certain circumstances (particularly, if the gauge coupling
constant between gravitino and vector fields in the
supergravity supermultiplet is non-zero)
the gauge symmetry principle may require non-zero cosmological constant.

In April 1978, when the original enthusiasm
on the possibility to build a complete fundamental
theory in the framework of supergravity changed to skepticism,
we arrived at the idea to generalize supergravity to higher spins
(in particular, to spin 5/2).
It is well known that massless fields of spin 1 $(A_n)$, 3/2
$(\psi_{n\ga})$  and spin 2 $(g_{nm})$ are gauge fields with the
transformation laws
\be
\delta A_n =\partial_n \varepsilon\,,\qquad
\delta \psi_{n\ga}= \partial_n \varepsilon_\ga \,,\qquad
\delta g_{nm}= \partial_n \varepsilon_m + \partial_m \varepsilon_n \,.
\ee
($m,n \ldots $ are vector indices and $\alpha ,\beta \ldots $ are
spinor indices).
The situation with HS massless fields is analogous.
As was first shown by Fronsdal \cite{Fron}, an
integer-spin massless spin$-s$ gauge field is described by the totally
symmetric tensor $\varphi _{n _1\ldots n_s}$ subject to the double
tracelessness condition \cite{Fron} $\varphi ^{r}{}_{r}{}^{s}{}_{s n
_5\ldots n_s}=0$ which is nontrivial for $s\geq 4$. A quadratic action
\cite{Fron} for a free spin $s$ field $\varphi _{n _1\ldots n_s}$
is fixed unambiguously \cite{C} (up to an overall factor) in the form
$S_s = \varphi L \varphi $ with some second order differential operator
$L$ by the requirement of gauge invariance
under the Abelian gauge transformations
\begin{equation}
\delta\varphi_{n_1\ldots n_s}=\partial_{\{n_1}\varepsilon_{n_2\ldots n_{s}\}}
\end{equation}
with the parameters $\varepsilon_{n _1...n _{s-1}}$ which are rank-$(s-1)$
totally symmetric traceless tensors,
$\varepsilon^{r}{}_{r n_3\ldots n_{s-1}}=0$.
This formulation is parallel \cite{WF} to the
metric formulation of gravity and is called
formalism of symmetric tensors.
Fermionic HS gauge
fields are described analogously \cite{FFron}
in terms of rank-$(s-1/2)$ totally
symmetric spinor-tensors $\psi _{n _1...n _{s-1/2}\alpha }$
subject to the $\gamma -$tracelessness condition $\gamma
^s {}_{\alpha }  {}^{\beta } \psi ^r{} _{r s n _4...n
_{s-1/2}\beta }=0$.

The problem was to introduce interactions of HS fields
with some other fields in a way compatible with HS gauge
symmetries. A particularly important example is provided
by the interaction with gravity.
It was straightforward to see that the standard covariantization
procedure $\partial \rightarrow D=\partial -\Gamma $
breaks down the invariance under the HS gauge
transformations because it turns out that, in order to prove invariance of the
action $S_s$, one has to commute derivatives, while the commutator of
the covariant derivatives is proportional to the Riemann tensor,
$[D\ldots,\, D\ldots]={\cal R}\ldots\,\, $.
As a result, the gauge variation of the
covariantized action $S_s^{cov}$ has the following structure:
\begin{equation}
\label{fvar}
\delta S_s^{cov} ={\cal R}_{\ldots}(\varepsilon_{\ldots}
D\varphi_{\ldots})\neq 0\,.
\end{equation}
It was not clear what to do with these terms because for $s >2$
they contain the Weyl part of the Riemann tensor
that cannot be compensated by any
transformation of the gravitational field \cite{diff}.
However, having learnt the role of the cosmological
constant in the gauged supergravity, it was natural to reconsider
the problem in the AdS background. (As a result,
our study was not affected by the papers on the
difficulties
of the gravitational interaction of HS fields \cite{diff},
all assuming implicitly an expansion near the flat space.)
Somewhat later this led to the
desired result as was originally checked for spin 3
by a rather complicated straightforward analysis \cite{unp}.
A number of interesting papers
on the existence of some consistent interactions of the HS
fields appeared in eighties both in the light-cone approach
\cite{LC} and in the covariant approach
\cite{cov}, providing a strong
evidence that some fundamental HS gauge theory must exist.
However, in all these works, the problem was
considered in the flat space and therefore no progress with the
gravitational interaction was achieved.

The key point of the relevance of the
AdS background is as follows.
Near the AdS background, the Riemann tensor ${\cal R}$ is not small but
${\cal R} = R +\lambda^2 g g$,
where $\lambda^{-1}$ is the radius of the AdS space, $g$ is the
background AdS metric tensor and ${ R}$ denotes a
fluctuation of the Riemann tensor near the background curvature.
When expanding around the AdS geometry one therefore has
to expand in powers of ${ R}$ rather than in powers of the
Riemann tensor itself. In other words, in the AdS space
the commutator of covariant derivatives is of order $\lambda^2$.
Now one can modify the action by adding some
cubic terms of the form
\be
\label{cact}
S^{int} =\int_{M_4}\sum_{A,B}\lambda^{-(A+B)}
D^A (\phi ) D^B (\phi )
{ R}
\ee
containing higher derivatives of the dynamical fields with the
coefficients proportional to negative powers of $\lambda$.
Note that for any two given spins a highest order of the
derivatives in a vertex is finite increasing linearly with the sum of
spins. As we checked originally for the case of spin 3 \cite{unp},
there exists such a unique (modulo total derivatives and field
redefinitions) action (\ref{cact})  that its HS gauge variation
exactly compensates the original variation (\ref{fvar}).

The fundamental concept of the HS gauge
theory is the underlying symmetry principle. By
construction, the class of HS gauge theories consists of most
symmetric theories
having as many as possible symmetries unbroken (any more symmetric
theory will have more lower and/or higher spin
symmetries and therefore will belong to the class of HS theories).
As such HS gauge theory is of particular importance for
the search of a fundamental symmetric phase of the superstring
theory. This is most obvious in the context
of the so called Stueckelberg
symmetries in the string field theory which have a form of some spontaneously
broken HS gauge symmetries. Whatever a symmetric phase of the superstring
theory is, Stueckelberg symmetries are expected to become unbroken HS
symmetries in such a phase
and, therefore, the superstring field theory has to become
one or another version of the HS gauge theory.
One explanation why the HS gauge theory has not been yet
observed in the superstring theory is that still no complete
formulation of the latter is known in the AdS background.

To elucidate a structure of the HS algebra we used
the so-called ``geometric" approach to supergravity \cite{geom,MM}
in which fields, action and transformation laws are formulated in terms
of the gauge fields of some (super)algebra identified with the global
symmetry algebra of a theory. An attractive feature of this machinery
is that it avoids explicit contraction of indices by the metric tensor,
treating all fields as differential forms. The distinguishing property
of the metric tensor that appears
on equal footing with other massless fields in a HS multiplet
is that it has a nonzero vacuum expectation value
providing a meaningful linearized approximation for all fields
in the model.

As a first step towards such a formulation it was
instructive to check whether it is possible to reformulate
the free dynamics of the HS fields in terms of some
curvatures linear in the dynamical fields and containing the
background (Poincare or AdS) gravitational vierbein and Lorentz
connection 1-forms. As the lowest nontrivial dimension with propagating
HS fields is d=4, we first focused on this case.
It was shown that such free actions
do indeed exist
\cite{Fort1} (the generalization to any dimension was later given in
\cite{LV} thus indicating that the same approach is likely to
be successful in any $d$). The resulting formulation of the free field
HS dynamics is a generalization of the Cartan (frame) formulation of
gravity. Assuming that thus found linearized HS curvatures
result from the linearization of some non-Abelian HS
curvatures, one deduces from their form some of the structure coefficients of
the full HS algebra (namely those that correspond to the commutators of
any generator with the generators of the space-time algebra (Poincare
or AdS) giving rise to the gravitational gauge fields).
These data fix (up to a multiplicity) a possible pattern
of the reduction of the full algebra with respect to its space-time
subalgebra and can be used as a sort of ``initial data" for the
problem of reconstruction of the full non-Abelian HS algebra. Adding
some (more or less) natural assumptions on the structure of the
HS algebra, its structure coefficients were found explicitly
in \cite{FVa} by a direct solution of the Jacobi identities. A few
years later it was observed \cite{Fort2} that this
algebra admits a simple realization in terms of the oscillator
(i.e., star product) algebra with spinor generating elements.
As we demonstrate in this talk
this fact links together such seemingly
different properties of the HS theories as the relevance of the
AdS background, necessity of introducing infinitely many spins in the
HS models and space-time non-locality of the HS
interactions manifested by the formula (\ref{cact}). Note
that altogether these properties make the HS theories reminiscent
of the superstring theory with the analogy between the cosmological constant
and the parameter $\alpha^\prime$. Remarkably,
the properties of the HS theory that follow from the structure
of the HS algebra found in eighties \cite{FVa,Fort2}
``predicted" some of the present day hot topics in the
superstring theory.

\section{D=4 Higher Spin Algebra and $AdS$ Vacuum}
\label{Higher-Spin Symmetries}

Perturbative HS dynamics is formulated in terms of the ``superfield"
1-form $\go (Y|x)=dx^\un \go_\un (Y|x)$ and 0-form  $C (Y|x)$
that describe, respectively, the  HS gauge fields and
(HS) Weyl tensors along with matter fields. They depend
on the space-time coordinates
$x^\un$ and commuting Majorana spinors
$Y_\Omega = ( y_\ga , \bar{y}_\dga )$ ($\Omega ,\Lambda ... = 1\div 4$;
$\ga,\gb = 1,2; \dga , \dgb =1,2$). The difference with the standard
superfield approach is that the additional spinor coordinates
$Y_\Omega$ commute
and therefore the HS superfields contain infinite chains
of the space-time component fields in their expansion in
power series in the spinor variables
\be
\label{gsf}
\go (y,{\bar{y}}|x)=
\sum^\infty_ {n,m=0}
   \! \frac{1}{2i m!n!} dx^\ull
\go_\ull{}^{\gal_1\ldots\gal_n\,{\dga}_1 \ldots {\dga_m}} (x)\,
    y_{\gal_1}\ldots y_{\gal_n}\,
    {\bar{y}}_{\dga_1}\ldots \bar{y}_{\dga_m}\,,
\ee
\be
\label{csf}
C (y,{\bar{y}}|x)=
\sum^\infty_ {n,m=0}
   \! \frac{1}{2i m!n!} C^{\gal_1\ldots\gal_n\,
{\dga}_1 \ldots {\dga_m}} (x) \,
    y_{\gal_1}\ldots y_{\gal_n}\,
    {\bar{y}}_{\dga_1}\ldots \bar{y}_{\dga_m}\,.
\ee
Here the component fields carrying  (odd) even numbers of spinor indices
are required to be (anti)commuting\footnote{To make the
relationship between spin and statistics more
obvious it is convenient to assume that $\go$ and $C$ depend on an
additional Clifford element
$\psi$ ($\psi^2 =1$)  requiring the fields $\go (\psi,Y|x)$ and
$C(\psi , Y |x)$ to be even functions of the auxiliary variables, i.e.
$\go (\psi,Y|x)=\go (-\psi,-Y|x)$
and $ C (\psi,Y|x)=C (-\psi,-Y|x)$. Obviously,
$\psi$ merely labels fermions and can be
discarded as in (\ref{gsf}) and (\ref{csf}). This construction is
particularly useful however in the HS models
with extended supersymmetry resulting \cite{Fort2}
from introducing  a
number of Clifford elements $\psi^i$, $i=1\div N$.}.
All the component fields
$\go_\un{}^{\gal_1\ldots\gal_n\,{\dga}_1 \ldots {\dga_m}} (x)$
with $n+m = 2(s-1)$ and
$C^{\gal_1\ldots\gal_n\,{\dga}_1 \ldots {\dga_m}} (x)$
with $|n-m| = s$ are associated with the spin $s$. The dynamical
spin $s\geq 1$ fields are identified with
$\go_\un{}^{\gal_1\ldots\gal_n\,{\dga}_1 \ldots {\dga_m}} (x)$
at $|n-m|= 0$ for bosons and $|n-m|=1$ for fermions. The matter fields
with $s=0$ and $1/2$ are identified with
$C^{\gal_1\ldots\gal_n\,{\dga}_1 \ldots {\dga_m}} (x)$ at $n=m=0$
and $n+m = 1/2$, respectively. All other fields express via
higher derivatives of the dynamical fields by virtue of
appropriate constraints \cite{Fort1}.

The 1-forms $\go(Y|x)$ are gauge fields of the HS
algebra. The HS field strength
has the standard form $R^A=d\go^A +f_{BC}^A \go^B \wedge \go^C$
where $f_{BC}^A$ are the structure coefficients of the HS
algebra originally found in \cite{FVa}. As shown in \cite{Fort2}
these HS curvatures result from the star product construction
\be
\label{HScu}
R(Y|x)=d\go (Y|x)-(\go\wedge * \go)(Y|x)\,,
\ee
where the  space-time coordinates $x^\un$ are commuting
while the star product acts on the auxiliary coordinates $Y$
according to the rule
\bee
\label{wprod}
 (f*g)(Y)
   &=&\frac{1}{(2\pi)^{2}}\int d^{4}U d^{4} V\exp(iU_\Go V^\Go)
   f(Y+U)g(Y+V)\,\nn\\
& =&
e^{i\frac{\partial}{\partial Y^1_\Go}
\frac{\partial}{\partial Y^{2\Go}}} f(Y+Y^1 ) B(Y+Y^2 ) |_{Y^1 = Y^2 =0}\,,
\eee
where $U_\Go V^\Go = U_\Go V_\Gpo C^{\Go \Gpo}$ and
$C_{\Go\Gpo}$ is the 4d charge conjugation matrix
used to raise and lower spinor indices
$U^\Go = C^{\Go\Gpo}U_\Gpo$, $U_\Go =U^\Gpo C_{\Gpo\Go}$.
This star product defines the associative algebra $A_4$
(called Weyl algebra) with the
defining relation
$
Y_\Go *Y_\Gpo - Y_\Gpo * Y_\Go = 2iC_{\Go\Gpo}\,.
$
It describes
the product of Weyl ordered (i.e. totally symmetric)
polynomials of oscillators in terms of symbols of operators \cite{sym}.
This Weyl product law (called Moyal
bracket \cite{moyal} for commutators constructed from (\ref{wprod})) is
obviously nonlocal manifesting the ordinary
quantum-mechanical nonlocality.

The 0-form $C(Y|x)$ belongs to the twisted
adjoint representation with the covariant derivative
\be
\label{DC}
DC(Y|x) = dC(Y|x) -(\go * C)(Y|x) + (C*\tilde{\go} ) (Y|x) \,,
\ee
where$\quad\tilde{}\quad$is the chirality flipping
automorphism of the algebra
\be
\tilde{f} (y,\bar{y} ) = f(y, -\bar{y} ) \,.
\ee
Note that in the bosonic case with $f(-y, -\bar{y}) = f(y,\bar{y})$ there
is no difference between changing a sign of either $y$ or $\bar{y}$.
In the fermionic case the two cases are not equivalent and, for the complete
formulation, it is necessary to double the fields as discussed in
the section \ref{Nonlinear Higher-Spin Equations}.

The HS gauge transformations are
\bee
\label{gtrc}
\delta \go (Y|x)&=&d\epsilon (Y|x)- (\go  * \epsilon ) (Y|x) +
(\epsilon * \go ) (Y|x)\,,\nonumber\\
\delta C (Y|x)&=& (\epsilon * C ) (Y|x) -
(C  * \tilde{\epsilon} ) (Y|x) \,.
\eee
Since, as explained in the section \ref{Free Equations}),
higher components in the
expansions of $\go(Y|x)$ and $C(Y|x)$  in powers of $Y$
identify with the higher
derivatives of the dynamical HS fields by virtue of constraints,
the field transformation law effectively contains higher space-time
derivatives of the dynamical fields in the terms containing
higher components of the gauge parameters $\epsilon (Y|x)$.
Thus, the
quantum-mechanical nonlocality in the auxiliary spinor coordinates
$Y_\Omega$ induces the space-time derivatives of all orders
in the HS transformation laws.
This is in accordance with the fact
\cite{cur,gol} that a
 spin $s$ conserved current $T_{n_1 \ldots n_s} \sim$
$\partial_{n_1} \ldots \partial_{n_l}\bar{\phi}
\partial_{n_l +1}\ldots\partial_{n_s} \phi$
contains derivatives of order $s$.

A structure of the full nonlinear HS equations
of motion is such that
any solution $\go_0$  of the  zero-curvature equation
\be
\label{vacu}
d\go = \go* \wedge \go\,
\ee
solves the HS equations.
Locally, such a vacuum solution admits a pure gauge form
$\go_0 = - g^{-1}(Y|x)* d g(Y|x)$
with some invertible element $g(Y|x)$ of the
Weyl algebra, $g*g^{-1} = g^{-1} *g = I$.
It breaks the local HS symmetry to its stability  subalgebra
with the infinitesimal parameters $\epsilon_0 (Y|x)$
satisfying the equation
$D_0 \epsilon_0\equiv  d\epsilon_0 -w_0 * \epsilon_0
+\epsilon_0 * w_0 =0$ which solves as
$\epsilon_0 (Y|x) = g^{-1}(Y|x)* \epsilon_0 (Y)* g(Y|x)$.
In the HS theories no further symmetry breaking
is induced by the field equations, i.e.
$\epsilon_0 (Y)$ parametrizes the global symmetry
of the theory.
Therefore, the HS global symmetry algebra
identifies with the Lie superalgebra constructed from the
(anti)commutators of the elements of the Weyl algebra.
Note that the
fields carrying odd numbers of spinor indices are
anticommuting thus inducing the superalgebra structure into
(\ref{vacu}).

Functions bilinear in $Y_\Omega$ form a
subalgebra with respect to star-commutators. This is
the $AdS_4$ algebra $sp(4;R)\sim o(3,2)$.
One can look for a solution of the vacuum
equation (\ref{vacu}) in the form
\be
\label{anz0}
\go_0 =
\frac{1}{4i}
\left ( \go_0^{\ga\gb} (x)y_\ga y_\gb
+ \bar{\go}_0^{\dot{\ga}\dot{\gb}} (x)\bar{y}_{\dot{\ga}}
\bar{y}_{\dot{\gb}}
+2\lambda h_0^{\ga\dot{\gb}} (x) y_\ga
\bar{y}_{\dot{\gb}}  \right )\,.
\ee
Inserting this formula into
(\ref{vacu}) one finds that the fields $\go_0$, $\bar{\go_0 }$ and
$h_0$ identify, respectively,
with the Lorentz connection and the frame field
of $AdS_4$ provided that the frame $h_0$ is
invertible. The parameter $\lambda= r^{-1}$ is identified with
the inverse AdS radius. Thus,  the star product origin of the HS algebra
leads to the AdS geometry as a natural vacuum solution.

Note that spin $s>2$ gauge fields are described by the degree $2(s-1)$
polynomials which do not close to any finite-dimensional subalgebra.
Therefore,  HS ($s > 2 $)  massless fields can only appear in
infinite sets of massless fields with infinitely increasing spins.
In other words, any finite-dimensional subalgebra of the HS symmetry
algebra containing
the Lorentz subalgebra can have at most spin 2 gauge fields.
One can speculate on some spontaneous
breakdown of the HS symmetries
down to a finite-dimensional subalgebra (followed by a flat
contraction via a shift of the vacuum energy in the
broken phase). In a physical phase with
$\lambda \sim 0$ and $m\gg m^{exp} $ for HS fields,
only usual sets of lower spin gauge fields can remain massless.

\section{Higher Spin Action in $AdS_4$}
\label{Higher Spin action in $AdS_4$}

The HS action used in \cite{FV1}
is formulated in terms of the components of
the curvatures (\ref{HScu})
\be
\label{rsf}
R (y,{\bar{y}}|x)=
\sum^\infty_ {n,m=0}
   \! \frac{1}{2i m!n!} R^{\gal_1\ldots\gal_n\,
{\dga}_1 \ldots {\dga_m}} (x)
    y_{\gal_1}\ldots y_{\gal_n}\,
    {\bar{y}}_{\dga_1}\ldots \bar{y}_{\dga_m}\,
\ee
as follows
\begin{equation}
\label{act31}
S=-\frac{1}{4\kappa^2 \lambda^2}
\sum_{m,n =0}^{\infty}
\frac{i^{n+m-1}}{n!m!} \epsilon
(n-m) \int_{M^4}  R_{\alpha_1\ldots\alpha_n \,\dot{\beta}_1\ldots\dot{\beta}_m} \wedge
R^{\alpha_1\ldots\alpha_n}\,{}^{\dot{\beta}_1\ldots\dot{\beta_m}}\,,
\end{equation}
where  $\epsilon (n)$ is a sign function, i.e.
$\epsilon (0)=0$ and $\epsilon (\pm n ) = \pm 1$ for any positive integer
$n$. To analyze this action perturbatively one expands the gauge fields
as $\go = \go_0 + \go_1$,
where $\go_0$ is a zero curvature vacuum solution (i.e. $R_0 =0$)
and $\go_1$ is the dynamical part. In what follows we identify
a vacuum solution with some $AdS_4$ fields of the form (\ref{anz0}).
The HS curvatures then expand as $R= R_1 +R_2$, where
\be
\label{exp2}
R_1 (Y|x) = d\go_1 (Y|x) -(\go_0 *\wedge \go_1 )(Y|x)
-(\go_1 *\wedge{\go}_0 ) (Y|x) \,.
\ee

The quadratic part of the action has the form (\ref{act31})
with $R_1$ instead of $R$. It was originally
found in \cite{Fort1} along with the form of the
linearized curvatures from the condition that its
variation with respect to the ``extra fields"
$\omega_{\alpha_1 \ldots\alpha_n\,\dot{\beta}_1 \ldots\dot{\beta}_m}$
with $|n-m|>2$ vanishes identically.  The variation with respect to
the dynamical
fields, $\omega_{\alpha_1 \ldots\alpha_n\,\dot{\beta}_1 \ldots\dot{\beta}_m}$
with $|n-m|\leq 2$ remains non-trivial and leads to the correct free
equations for a spin $s\geq 3/2$ massless field in the sector of
$n+m =2(s-1)$.

Beyond the quadratic order, the  extra fields do contribute
into the interaction terms. One therefore has to express them in terms
of the dynamical fields to have a well-defined nonlinear action.
The appropriate linearized constraints have the form \cite{Fort1}
\be
\label{con1}
h_{\{\alpha} {}^{\dot{\gamma}}\wedge R_{1\,\alpha_1 \ldots\alpha_n\}\,
\dot{\beta_1}\ldots\dot{\beta_m}\dot{\gamma}}=0\qquad (n> m \geq 0)\,
\ee
(and complex conjugate).
These constraints express algebraically all extra fields
$\omega_{\alpha_1 \ldots\alpha_n\,\dot{\beta}_1 \ldots
\dot{\beta}_m}$ with $|n-m|>2$
in terms of the dynamical fields
$\omega_{\alpha_1 \ldots\alpha_n\,\dot{\beta}_1 \ldots
\dot{\beta}_m}$ with $|n-m|\leq
2$ and their derivatives. They play a key role in the
description of the HS
dynamics, governing a form of the interactions of the
dynamical HS fields in the
action (\ref{act31}) and, in particular, giving rise to
higher derivatives and
negative powers of the cosmological constant in the HS interactions.

Remarkably, the action (\ref{act31}) supplemented with the
constraints (\ref{con1}) turns out to be gauge invariant
under (appropriately deformed) HS symmetries \cite{FV1}
up to $o(\phi^2) \epsilon$ terms (note that this requirement
fixes uniquely the relative coefficients in front of the individual
spin-$s$ quadratic actions in (\ref{act31})). This action is
explicitly general coordinate invariant
and reduces to the Einstein-Hilbert
action in the MacDowell-Mansouri form \cite{MM}
in the spin$-2$ sector $(n+m=2)$. Therefore, it solves
the original problem of a higher-spin-gravitational interaction at the
cubic order. Let us note that the described approach leads to
a particular form of the action (\ref{cact}) but
avoids complicated computations and is uncomparably
simpler than the straightforward analysis \cite{unp}.

A non-trivial question is how to extend
this result to the highest orders in interactions.
To proceed one has first of all to determine the full spectrum of
fields in the theory\footnote{Note that a
reduction of a full consistent action by setting
 any subset of fields  equal to zero still leads to an
 action  consistent at the cubic order
(as is most obvious from its Noether current character) even in case
this reduction cannot be consistently extended beyond the cubic
approximation.}.
Another problem is to
find an appropriate generalization of the linearized constraints
(\ref{con1}).
This program was completed at the level of the equations of motion
\cite{more} (see also \cite{gol}) in terms of
appropriate generating functions of the auxiliary spinor variables
satisfying certain flatness conditions (this formalism is referred
to as ``unfolded formulation").
As a result, the full answer to the  question on the true spectra
of HS and lower spin fields compatible with the consistent
interactions was given
and the full nonlinear system of constraints and
equations of motion was formulated. These results are summarized
in the section
\ref{Nonlinear Higher-Spin Equations}.

Let us mention that an alternative fruitful approach towards
HS gauge theory into which Fradkin and Metsaev
contributed a lot \cite{FLC} is the light-cone analysis of the
problem initiated in
\cite{LC}. It is complementary to the covariant approach: the latter
makes all symmetries explicit by virtue of introducing infinite sets of
auxiliary fields, while the former breaks down all HS gauge
symmetries to formulate dynamics directly in terms of the
dynamical degrees of freedom.
Another important direction in the HS gauge theory developed by
Fradkin and Linetsky is conformal HS gauge theory \cite{FLA,FL}.
Although, analogously to the case of conformal (super)gravity
(for more details and references on conformal supergravity
see \cite{CFT}),
it is not unitary in $d=4$ (i.e., contains ghosts)
conformal HS gauge theory can lead to a compact formulation of the unitary
(AdS) HS theory by virtue of introducing compensators. In the section
\ref{Towards Unfolded Formulation of the Conformal Higher Spin Theory}
we will announce some new result
on the unfolded formulation  of the linearized
conformal 4d HS theories that, hopefully,  provides a
good starting point to study a full nonlinear conformal HS gauge theory.

\section{Unfolded Form of the Free HS Equations}
\label{Free Equations}

{}From the constraints (\ref{con1})
along with the free equations of motion that follow from the
linearized action (\ref{act31}) one proves  by virtue of the
Bianchi identities \cite{Fort1} that
\be
\label{R_1}
R_1 (y,\bar{y}|x) = h^{\gga\dgb} \wedge h_{\gga}{}^{\dga}
\frac{\partial}{\partial \bar{y}^\dga}
\frac{\partial}{\partial \bar{y}^\dgb}
C(0 ,\bar{y} |x) +
 h^{\ga\dot{\gamma}} \wedge h^\gb{}_{\dot{\gamma}}
\frac{\partial}{\partial {y}^\ga}
\frac{\partial}{\partial {y}^\gb}
C({y},0 |x)\,,
\ee
i.e. most of the components of the HS curvatures vanish
on-mass-shell except for those
parametrized by the  holomorphic ($C({y},0 |x)$) and antiholomorphic
($C(0 ,\bar{y} |x)$) parts of $C(Y|x)$.
For spin 2 these are order--4 polynomials
in $y$ or $\bar{y}$ parametrizing the
Weyl tensor. For higher spins, $C({y},0
|x)$ and $C(0 ,\bar{y} |x)$ parametrize the ``HS Weyl tensors".
The equation (\ref{R_1}) is an alternative form of the
free HS field equations
equivalent to the standard one.

By virtue of the Bianchi
identities, the equation (\ref{R_1}) imposes some differential
conditions on the Weyl tensors. As shown in \cite{Ann},
for spins $s\geq 3/2$ these differential conditions turn out
to be equivalent to the equation
\be
\label{cald0}
{D}_0 C \equiv d C -{w}_0 *C +C* \tilde{w}_0  = 0 \,,
\ee
where $D_0$ is the linearized covariant derivative  (\ref{DC}).
For the $s=1$ case, the equation (\ref{R_1}) does not impose
any restrictions on the
holomorphic and antiholomorphic 0-forms
$C(Y|x)$ bilinear in $Y$ that
parametrize the spin 1 field strength. The Maxwell equation
is contained in (\ref{cald0}). For the cases of spin 0 and spin 1/2
matter fields there are no associated gauge fields and the respective
(Klein-Gordon and Dirac) equations are also contained in (\ref{cald0}).
To summarize, the content
of the equations (\ref{R_1}), (\ref{cald0})
is twofold: they describe usual free field equations
for all spins and express all extra
gauge fields and 0-forms $C(Y|x)$ in terms of the
higher derivatives of the dynamical fields.

Since $\tilde{} $ changes a sign of the $AdS_4$
translations thus transforming
the commutator in the adjoint representation into the anticommutator
in (\ref{DC}) for the terms linear in $h$,
$D_0$ in (\ref{cald0}) has the form
$D_0 C (y,{\bar{y}}|x) \equiv
D^L C (y,{\bar{y}}|x)
+\frac{i\lambda}{2}\{ h^{\ga\dgb} y_\ga \bar{y}_\dgb , C (y,{\bar{y}}|x) \}_* $
where $\{a,b\}_* = a*b +b*a$ and
$D^L$ is the Lorentz covariant derivative,
\be
\label{dlor}
D^L C (y,{\bar{y}}|x)=
d C (y,{\bar{y}}|x) +\frac{i}{4}
([\go^{\ga\gb}y_\ga y_\gb +\bar{\go}^{\dga\dgb}\bar{y}_\dga \bar{y}_\dgb ,
C (y,{\bar{y}}|x)  ]_* )\,.
\ee
By virtue of (\ref{wprod}) we get
\be
\label{LL}
D_0 C (y,{\bar{y}}|x) \equiv
D^L C (y,{\bar{y}}|x) +i\lambda h^{\ga\dgb}
\Big (y_\ga \bar{y}_\dgb -\frac{\partial}{\partial y^\ga}
\frac{\partial}{\partial \bar{y}^\dgb}\Big ) C (y,{\bar{y}}|x) =0\,.
\ee
{}From this expression it is clear that
(\ref{cald0}) links the derivatives in the
space-time coordinates $x^\un$ with those in the
auxiliary spinor variables $y_\ga$ and $\by_\dga$.
Schematically, in the sector of
0-forms $C(Y|x)$, the derivatives in the auxiliary spinor variables
form a square root of the space-time derivatives,
$
\frac{\partial}{\partial x^\un } C(y,\by |x)
\sim \lambda  h_\un{}^{\ga\dgb}
\frac{\partial}{\partial y^\ga}
\frac{\partial}{\partial \by^\dgb} C(y,\by |x)\,.
$
Explicit component expression is
\be
\label{Cn>m}
\!\!C_{\ga_1 \ldots \ga_n\,,\dgb_1\ldots \dgb_m }\!=\!
\frac{1}{(2i\lambda)^m} h^{\ull_1}_{\ga_1\dgb_2}  D^L_{\ull_1}
\!\ldots h^{\ull_m}_{\ga_{m}
\dgb_{m}}\!
D^L_{\ull_m}\!
C_{\ga_{m+1}\ldots  \ga_n}\,\quad n \geq m
\ee
(and complex conjugated). Analogous formulas
are true for the gauge 1-forms \cite{Fort1}. Note that these expressions, as
well as those for the extra gauge fields resulting from the resolution of the
constraints (\ref{con1}), contain  negative powers of  $\lambda$.
As a result, insertion of the expressions for the extra fields in terms of the
dynamical HS fields into the action (\ref{act31}) gives rise to some terms
containing higher derivatives and negative powers of $\lambda$ in the
interaction terms. Note also that the requirement that the free action is
independent of the extra fields acquires now a simple interpretation as the
condition that the free action has to be free of higher derivatives and
regular in the flat limit.

A few comments are now in order.

\noindent
(i) AdS vacuum geometry implies that the vacuum fields (\ref{anz0})
   are bilinear in the spinor oscillators and, as a consequence,
   that the relationship (\ref{LL}) is local, containing at most
   two derivatives in $Y$. This guarantees that the free field HS dynamics
   in $AdS_4$ is local.

\noindent
(ii) The nonlocality in $Y$ at the interaction level
    induced by the star product (\ref{wprod}) implies the space-time
    nonlocality of HS interactions.

\noindent
(iii) As a consequence of the field equations,
the gauge transformation law (\ref{gtrc})
contains effectively the  higher space-time
    derivatives via higher derivatives in $Y$ induced by the star
    product. The higher power in $Y$ (i.e., spin) of the HS
   gauge parameter is, the higher space-time derivatives appear
   in the transformation law.

\noindent
(iv) Vacuum solutions in the HS theory different from the
     most symmetric $AdS_4$ one may
     give rise to nonlocality in the linearized approximation.
    An interesting problem for the future is to study HS theory
    in the magnetic background to see whether or not it leads to
    the same type of nonlocality as in the non-commutative regime
   in the string theory \cite{mmoyal}.

Thus, the following facts turn out to be strongly correlated:
(i) HS algebras are described by the
star product in the auxiliary spinor space;
(ii)
relevance of the AdS background and
(iii) potential space-time nonlocality of the HS interactions
     due to the appearance of higher derivatives at the nonlinear level.

\section{Nonlinear Higher Spin Equations}
\label{Nonlinear Higher-Spin Equations}

Now we discuss the full nonlinear system of  4d HS equations
following to \cite{more,gol}.
The key element of the construction consists of the
extension of the space of auxiliary variables by the
doubling of auxiliary Majorana spinor variables $Y_\Go$
in the HS 1-form
$\go(Y|x)\longrightarrow W(Z;Y;K|x)$
and 0-form
$C(Y|x)\longrightarrow B(Z;Y;K|x)$ and by introducing a pair of
Klein elements $K=(k,\bar{k})$ having the properties
$
k^2 = \bar{k}^2 =1 \,,\quad k\bar{k} = \bar{k} k\,,
$
\bee
k f(z_\ga ,\bar{z}_\dga ;y_\ga , \bar{y}_\dga) &=&
f(-z_\ga ,\bar{z}_\dga ;-y_\ga , \bar{y}_\dga )k \,,\nn\\
\bar{k} f(z_\ga ,\bar{z}_\dga ;y_\ga , \bar{y}_\dga) &=&
f(z_\ga ,-\bar{z}_\dga ;y_\ga ,- \bar{y}_\dga )\bar{k} \,.
\eee
By definition, $\bar{k}$ generates the
automorphism$\quad\tilde{}\quad$while
$k$ generates the conjugated automorphism. The dependence on $K$
leads to the doubling of the HS fields necessary in presence of fermions
and also gives rise to some auxiliary fields \cite{Ann,gol}.

The dependence on the additional variables $Z_\Go$ is determined
in terms of the ``initial data" identified with the HS fields discussed
so far
\be
\label{inda}
W(0;Y;k,\bar{k}|x)=\go (Y;k,\bar{k}|x)\,,\qquad
B(0;Y;k,\bar{k}|x)=C(Y;k,\bar{k}|x)\,
\ee
by appropriate equations formulated below and effectively
describes all nonlinear corrections to the HS field equations.
It is convenient to introduce a new compensator-\-type spinor field
$S_\Go (Z;Y;Q|x)$ which does not carry its own degrees of freedom.
It plays a role of a covariant differential along the
additional $Z_\Go$ directions. To interpret $S_\Go (Z;Y;K|x)$ as
a $Z-$ 1-form $S=dZ^\Go S_\Go$ we introduce the
anticommuting $Z-$differentials $dZ^\Go dZ^\Gl=-dZ^\Gl dZ^\Go$.

The nonlinear HS dynamics is formulated in terms of the associative
star product
\be
\label{star2}
\!\!(f*g)(Z;Y)=\frac{1}{(2\pi)^{4}}
\int d^{4} U\,d^{4} V \exp{[iU^\Gl V^\Go C_{\Gl\Go}]}\, f(Z+U;Y+U)
g(Z-V;Y+V) ,
\ee
where $ U^\Gl $ and $ V^\Gl $ are spinor integration variables.
The star product (\ref{star2})  again yields a particular
realization of the Weyl algebra
corresponding to the normal ordering with respect to the
creation and annihilation operators
 $  B^\Gl = \frac{1}{2} (Y^\Gl - Z^\Gl )$ and
   $A_\Gl = \frac{1}{2} (Y_\Gl + Z_\Gl )$
obeying the commutation relations
\be
\label{ccom}
 [A_\Gl, A_\Go]_*=[B^\Gl, B^\Go]_* =0 \,,\qquad
   [A_\Gl, B^\Go]_* =i\delta_\Gl^\Go \,.
\ee
The following simple formulas are true
\be
\label{z,f}
    [Y_{\Gl}, f]_*=2i{\partial f\over \partial Y^\Gl}\,,\qquad
    [Z_{\Gl}, f]_*=-2i{\partial f\over \partial Z^\Gl} \,
\ee
 for any $f(Z,Y)$.
{}From (\ref{star2}) it follows that functions $f(Y)$ independent
of $Z$ form a proper subalgebra with the Weyl star product
(\ref{wprod}).

The full system of 4d equations has the form
\be
\label{dW}
dW=W*W\,,\qquad
dB=W*B-B*W\,,\qquad
dS=W*S-S*W\,,
\ee
\be
\label{SS}
S*B=B*S\,, \qquad
S*S= dZ^\Go dZ^\Gl \,(-iC_{\Go\Gl}+4 R_{\Go\Gl}(B))\,.
\ee
The function $R_{\Go\Gl}(B) $ that encodes all information
about the HS dynamics has the form
\be
\label{Rab}
 dZ^\Go dZ^\Gl \, R_{\Go\Gl}(B)=\frac{1}{4i} \Big (
dz_\alpha dz^\alpha \,{\cal F}({B})*( k e^{iz_\ga y^\ga })
+ d\bar{z}_{\dot{\alpha}}\, d\bar{z}^{\dot{\alpha}}\,
\bar{{\cal F}}({B})
*( \bar{k} e^{i\bar{z}_\dga \bar{y}^\dga}) \Big )\,,
\ee
where ${\cal F}(B)$ is some star
product power series in $B$ that parametrizes an ambiguity in
the HS interactions.  Even the simplest choice
${\cal F}({B}) = {B}$ leads to the nontrivial (nonlinear) dynamics.
The case ${\cal F}=0$ leads to the free field equations.
Let us note that the equations (\ref{SS}) have the
form of the deformed oscillator algebra \cite{DO}
equivalent to what is sometimes referred in the literature as
fuzzy sphere \cite{fs}.

The equations (\ref{dW}) and (\ref{SS})
are invariant under the gauge transformations
\be
\label{deltaW}
      \delta W=d\gvep+[\gvep , W ]_* \,,\qquad
      \delta S=[\gvep , S ]_*  \,,\qquad
    \delta B=[\gvep ,B]_* \,.
\ee
The space-time differential $d$ only emerges in the
 equations (\ref{dW}) which have a form of zero-curvature and
covariant constancy conditions and therefore admit explicit
solution in the pure gauge form
\be
\label{PG}
W = -g^{-1}(Z;Y;Q|x)* d g(Z;Y;Q|x)\,,
\ee
\be
\label{GB}
B (Z;Y;Q|x) = g^{-1}(Z;Y;Q|x)* b (Z;Y;Q)* g(Z;Y;Q|x) \,,
\ee
\be
\label{GS}
S (Z;Y;Q|x) = g^{-1}(Z;Y;Q|x)*s (Z;Y;Q)* g(Z;Y;Q|x)
\ee
with some invertible $g(Z;Y;Q|x)$ and arbitrary $x-$independent
functions \hfil\\ $b (Z;Y;Q)$ and $s (Z;Y;Q)$. Due to the gauge invariance
of the whole system one is left with only the equations
(\ref{SS})
for $b (Z;Y;Q)$ and $s(Z;Y;Q)$. These encode in
a coordinate independent way all information about the dynamics
of massless fields of all spins. In fact, the ``constraints"
(\ref{SS}) just impose appropriate restrictions on
$b$ and $s$ to guarantee that the original
space-time equations of motion are satisfied.  Let us stress
some parallelism between this coordinate-free formulation of the
HS dynamics and Matrix formulation of the suprestring theory. {}From
this perspective the equations (\ref{SS}) provide a covariant
``matrix" formulation of the HS gauge theory.

A simplest vacuum solution of the equation
(\ref{SS}) is $B_0 = 0$ and
$S_0 = dZ^\Go Z_\Go$.
{}From (\ref{z,f}) it follows
that
\be
\label{S0f}
[S_0, f]_* =-2i\partial f\,,\qquad
\partial =dZ^\Go \frac{\partial}{\partial Z^\Go }\,.
\ee
Interpreting the deviation of the full field $S$ from
the vacuum value $S_0$ as a $Z-$component of the gauge field,
$S=S_0 +2i dZ^\Go W_\Go$,
one rewrites the equations (\ref{dW}), (\ref{SS}) as
\be
\label{calr}
{\cal R} =
 dZ^\Go dZ^\Gl \,R_{\Go\Gl} (B)\,,\qquad
{\cal D} B =0\,,
\ee
where the generalized curvatures and covariant derivative
are defined by the relations
\be
\label{defcalr}
{\cal R} = (d+\partial ) (dx^\un W_\un +dZ^\Go W_\Go )
-(dx^\un W_\un +dZ^\Go W_\Go )\wedge
(dx^\un W_\un +dZ^\Go W_\Go )\,,
\ee
\be
\label{defcald}
{\cal D} ( A) = (d+\partial ) A-
(dx^\un W_\un +dZ^\Go W_\Go ) * A +
A* (dx^\un W_\un +dZ^\Go W_\Go ) \,.
\ee
($dx^\un dZ^\Go =-dZ^\Go dx^\un$.)
We see that the functions $R_{\Go\Gl} (B)$ in (\ref{SS})
identify with the $ZZ$ components of the generalized
curvatures, while $xx$ and $xZ$ components of the
curvature vanish. The equation  ${\cal D} B =0$ means
that the curvature $R_{\Go\Gl} (B)$ is covariantly constant.
In fact, it is the compatibility condition
for the equations (\ref{SS}), (\ref{Rab}) and
(\ref{defcalr}).

The consistency of the system of
equations (\ref{dW}), (\ref{SS}) guarantees that it admits a
perturbative solution as a system of differential equations
with respect to $Z_\Go$. A natural vacuum solution is
$
W_0 (Z;Y;k,\bar{k}|x) = \go_0 (Y|x),$ $ B_0 =0$ and
$ S_{0\Go} = Z_\Go$
with the field $\go_0$ (\ref{anz0}) describing the $AdS_4$ vacuum.
All fluctuations of the fields can be
expressed modulo gauge transformations in terms of the
initial data (\ref{inda})  identified with the physical HS
fields. Inserting thus obtained
expressions into (\ref{dW}) one reconstructs all
nonlinear corrections to the free field equations (\ref{R_1}),
(\ref{cald0}).

\section{Towards Unfolded Formulation of the 4d Conformal Higher Spin Theory}
\label{Towards Unfolded Formulation of the Conformal Higher Spin Theory}

The results discussed so far were established quite some time ago.
Now we announce some recent development in the 4d conformal HS theory
originally proposed by Fradkin and Linetsky \cite{FLA,FL}.

The extension to the conformal case is based on the embedding
$o(3,2) \sim sp(4|R) \subset su(2,2) \sim o(4,2)$. The conformal
generators can be realized as the bilinears built from the
bosonic oscillators $A_\Go = (a_\ga , \bar{a}_\dga ) $ and
$B^\Go = (b^\ga , \bar{b}^\dga )$ satisfying the commutation relations
(\ref{ccom}). The particle number operator $N=-i\{A_\Go, B^\Go\} $
identifies with the central element of $u(2,2)$.
The $4d$ conformal HS algebra
\hsa  was defined by Fradkin and Linetsky \cite{FLA} as the subalgebra
of the (Lie superalgebra built from the) Weyl algebra spanned by the
elements having equal numbers of the oscillators $A_\Go$ and $B^\Go$,
i.e. \hsa is the centralizer of $N$. The HS curvatures
have the form (\ref{HScu}) with the gauge fields
$\go = \go (A;B |x )$ and the appropriately modified star product.

Fradkin and Linetsky have shown \cite{FL} that there exists a
generalization of the $AdS_4$ action (\ref{act31}) to the conformal
case, which is bilinear in the conformal HS curvatures and
gauge invariant at the cubic level. At the free level it contains
higher derivatives (the higher spin is the higher derivatives appear) and
 describes non-unitary dynamics (contains ghosts).  Nevertheless,
the conformal HS theory may be useful to describe the unitary
$AdS_4$ HS theory with the help of compensators. It is therefore
interesting to study whether the ``unfolded" machinery
originally developed for the $AdS_4$ case can be applied to the
conformal HS theory. Here we make a first modest step in this
direction formulating unfolded form for the constraints for
the extra fields in the 4d conformal HS gauge theory.

Let us fix the background 1-form gauge field in the form
$\go_0 = h^\ga{}_\dgb a_\ga \bar{b}^\dgb$,
where  $h^\ga{}_\dgb$ is the flat Minkowski frame
identified with the sigma matrices.
(Note that the conformal HS theory admits the flat background
not only at the quadratic level but also at the interaction level \cite{FL}.)
Since $h^\ga{}_\dgb$ can be chosen $x$-independent and $\go_0$
 is built from mutually commuting oscillators,
the zero curvature equation (\ref{vacu}) is trivially satisfied.

As in the $AdS_4$ case, the component conformal HS fields
 in the expansion of the gauge field $\go$ in powers of the
auxiliary spinor variables classify either as dynamical fields
satisfying the conformal field equations
or as auxiliary
and extra fields that express algebraically in terms
of the dynamical fields and their derivatives by virtue of some
constraints. The
dynamical fields are identified with those carrying the lowest conformal
dimension with respect to the dilatation generator
$D=-i(a_\ga b^\ga-\bar{a}_\dga \bar{b^\dga}) $.
Fradkin and Linetsky
have suggested the constraints analogous to (\ref{con1})
 that express all fields in terms of the dynamical fields
and their derivatives.
With the aid of these constraints they have shown that the
conformal HS action remains gauge invariant in the first
nontrivial order in the interactions. A comment we want to make here
is that the constraints of the conformal HS gauge theory admit an
equivalent unfolded form
\bee
\label{RC}
R(A,B|x)&=&h^\ga{}_\dgb \wedge h^\gga{}^\dgb
\frac{\partial^2}{\partial b^\ga \partial b^\gga}
C(a,b,\bar{a} ,0|x)\nn\\
&+& h_{\gga \dga }\wedge h^\gga{}_\dgb
\frac{\partial^2}{\partial \bar{a}_\dga \partial \bar{a}_\dgb}
\bar{C}(0,b,\bar{a},\bar{b}|x)\,.
\eee
The compatibility conditions of (\ref{RC}) along with free  
conformal HS equations read
\bee
\label{cC}
d {C}(A,B|x)
&=& h^\ga{}_\dgb \Big ( \frac{\partial^2}{\partial b^\ga
\partial \bar{b}_\dgb}
- a_\ga \frac{\partial}{\partial \bar{a}_\dgb} \Big )
{C}(A,B|x ) \,,\nn\\
d \bar{C}(A,B|x)
&=& h^\ga{}_\dgb \Big ( \frac{\partial^2}{\partial a^\ga
\partial \bar{a}_\dgb}
- \bar{b}^\dgb \frac{\partial}{\partial {b}^\ga} \Big )
\bar{C}(A,B|x ) \,.
\eee
Here the fields associated with integer spins satisfy
$(N_a + N_{\bar{a}} - N_b - N_{\bar{b}}) \go(A,B|x) =0$
for gauge 1-forms and
$(N_a + N_{\bar{a}} - N_b + N_{\bar{b}}+2 )C(A,B|x) =0$,
$(N_a - N_{\bar{a}} + N_b + N_{\bar{b}}+2 )\bar{C}(A,B|x) =0$
 for Weyl 0-forms (here
$N_a=a_\ga\frac{\partial}{\partial a_\ga},$
$N_{\bar{a}}=\bar{a}_\dga \frac{\partial}{\partial \bar{a}_\dga} $,
$N_b = b^\ga\frac{\partial}{\partial b^\ga}$,
$N_{\bar{b}}= \bar{b}^\dga \frac{\partial}{\partial \bar{b}^\dga}$).
Note that the part of the equations (\ref{RC})  with negative conformal
dimension has the vanishing right hand side. Those with zero conformal
dimension contain a definition of the conformal HS Weyl tensors.

The equations (\ref{RC}) and (\ref{cC}) generalize the
$AdS_4$ equations (\ref{R_1}) and (\ref{cald0}) to the
conformal HS theory.
An important difference compared to the $AdS_4$ case is that
(\ref{RC})  describes the linearized
off-mass-shell constraints. A nonlinear
deformation of the equations (\ref{RC}) 
will therefore solve the problem of off-mass-shell
constraints in the conformal HS theory.
Hopefully, the development along these lines will lead to the
solution of the analogous problem in the $AdS_4$ HS theory via the
compensator mechanism, and, eventually, to a full Lagrangian formulation
of the both conformal and $AdS_4$ HS theories.
An elegant form of the equations (\ref{RC}) and (\ref{cC})
indicates a deep algebraic structure underlying the full conformal HS theory.

\section{Conclusions}

Let us summarize some key features of the HS
gauge theories.
HS gauge theories are based on the infinite-dimensional HS symmetries
\cite{FVa} realized as the algebras of
oscillators carrying spinorial representations of the space-time
symmetries \cite{Fort2}. These star product algebras exhibit
usual quantum-mechanical  nonlocality in the
auxiliary spinor spaces. The constraints in the HS gauge theory
transform this nonlocality into a
space-time nonlocality at the interaction level.
The same time the HS gauge theories remain local
at the linearized level because the space-time symmetries
are realized in terms  of bilinears in spinor oscillators.
The relevant geometric setting is provided by the Weyl bundle with space-time
base manifold and Weyl algebras with spinor generating elements as the fibre.
The star product acts in the fiber rather than directly in the space-time.
The noncommutative  Yang-Mills theory structure also appears in the fiber
sector. This is different from
the non-commutative Yang-Mills theory in the string
theory \cite{mmoyal}.
Another important difference is that
some of the ingredients of the theory  like twisted
adjoint representation used to describe matter fields and HS
Weyl tensors do not admit a deformation
quantization interpretation.
It is interesting to see whether these features
of the HS theory will get some counterparts in the string theory.

Finally, let us note that an important
analogy with the string theory is  that the HS theory is
based on the associative (star product) algebras and therefore
shares many of the features of the non-commutative geometry.
In particular, the whole construction extends
\cite{Ann} to the case with inner
symmetries by endowing all fields with the matrix indices.
HS gauge theories with
non-Abelian symmetries were classified in \cite{KV1}
(where also the systematic notation for the HS algebras was
introduced) in a way analogous to the Chan-Paton
symmetries for oriented and non-oriented strings.

\noindent{\bf Acknowledgments.}
This research was supported in part by
INTAS, Grant No.99-1-590 and by the RFBR Grants No.99-02-16207
and 00-15-96566.

\end{document}